\documentclass[runningheads]{llncs}
\usepackage[inkscapelatex=false]{svg}
\usepackage{multirow}
\usepackage{amsmath}
\usepackage{wrapfig}
\usepackage{tcolorbox}
\usepackage{booktabs}
\usepackage{array}
\usepackage{fontawesome5}
\usepackage{enumitem}

\usepackage[T1]{fontenc}
\usepackage{graphicx}
\usepackage{hyperref}

\begin{document}
%
%
\title{PriHA: A RAG-Enhanced LLM Framework for Primary Healthcare Assistant in Hong Kong}
\titlerunning{PriHA}

\author{
    Richard Wai Cheung Chan\inst{\dagger} \and
    Shanru Lin\inst{\dagger} \and
    Ya-nan Ma\inst{\ddagger} \and
    Hao Chen\inst{\mathparagraph} \and
    Liangjun Jiang\inst{\dagger(\textnormal{\faEnvelope})} \and
    Wenqi Fan\inst{\dagger(\textnormal{\faEnvelope})}
}
\authorrunning{R. W. C. Chan et al.}
%
\institute{
    $^{\dagger}$ The Hong Kong Polytechnic University, Hong Kong, China \\
    $^{\ddagger}$ Haikou Affiliated Hospital of Central South University Xiangya School of Medicine, Haikou, China \\
    $^{\mathparagraph}$ Sun Yat-Sen University Cancer Center, Guangzhou, China \\
    \email{\{24032618g, shanru.lin\}@connect.polyu.hk, hnmayn0987@163.com, chenhao@sysucc.org.cn, \{jljchina2025, wenqifan03\}@gmail.com} \\
    Corresponding Authors: Liangjun Jiang and Wenqi Fan.
}
\maketitle 
\begin{abstract}
To address the unsustainable rise in public health expenditures, the Hong Kong SAR Government is shifting its strategic focus to primary healthcare and encouraging citizens to use community resources to self-manage their health. However, official clinical guidelines are fragmented across disparate departments and formats, creating significant access barriers. While general-purpose Large Language Models (LLMs) such as ChatGPT and DeepSeek offer potential solutions for information accessibility, they are prone to generating factually inaccurate content due to a lack of localized and domain-specific knowledge. To this end, we propose a Retrieval-Augmented Generation-Enhanced LLM system as \textbf{Pri}mary \textbf{H}ealthcare \textbf{A}ssistant (\textbf{PriHA}) in Hong Kong. 
Specifically, a tri-stage pipeline is proposed that leverages a query optimizer to generalize user intent-oriented sub-queries, followed by a novel Dual Retrieval Augmented Generation (DRAG) architecture for mixed-source retrieval and context-reorganized generation. 
Comprehensive experiments and a detailed case study demonstrate that our proposed method can outperform both ablations and baseline in terms of accuracy and clarity. Our research provides a reliable and traceable dialogue retrieval framework for exploring other high-risk, localized application scenarios.


\end{abstract}
\section{Introduction}
Hong Kong's population is aging rapidly. By 2046, the proportion of residents aged 65 and above is projected to exceed 36\%\footnote{Hong Kong Population Projections for 2022 to 2046, Census and Statistics Department:\url{https://www.censtatd.gov.hk/en/EIndexbySubject.html?pcode=B1120015&scode=190}}. This demographic shift is leading to a higher prevalence of chronic diseases, imposing a heavier burden on public hospital services and medical expenditures. Accordingly, “\textbf{Prevention is better than Cure}” has become the core policy to maintain the sustainability of the public health system in Hong Kong. This strategy aims to shift the focus from treatment-oriented secondary medical services to prevention-oriented primary healthcare\footnote{Primary Healthcare Blueprint:\url{https://www.primaryhealthcare.gov.hk/bp/sc/index.html}}, emphasizing the need for patients to develop a deeper understanding of their health status and undergo earlier screening. While the government has published extensive healthcare guidelines and disease-screening schemes, a critical, often-overlooked dimension is the information gap. A report indicates that over half of the participants feel unequipped to access the care guidance they need \cite{hkcss2023_exercise}. This divide is further exacerbated by the fragmentation of information across disparate government websites, where official guidelines are often stored in hard-to-retrieve formats, such as tables and figures within PDF files. 
Furthermore, certain key documents are available only in English, creating language barriers for the local Chinese-speaking population. To enable citizens to manage their health effectively, a more accessible information platform is urgently needed.

\begin{figure}
    \centering
    \vskip -0.25in
    \includegraphics[width=0.80\linewidth]{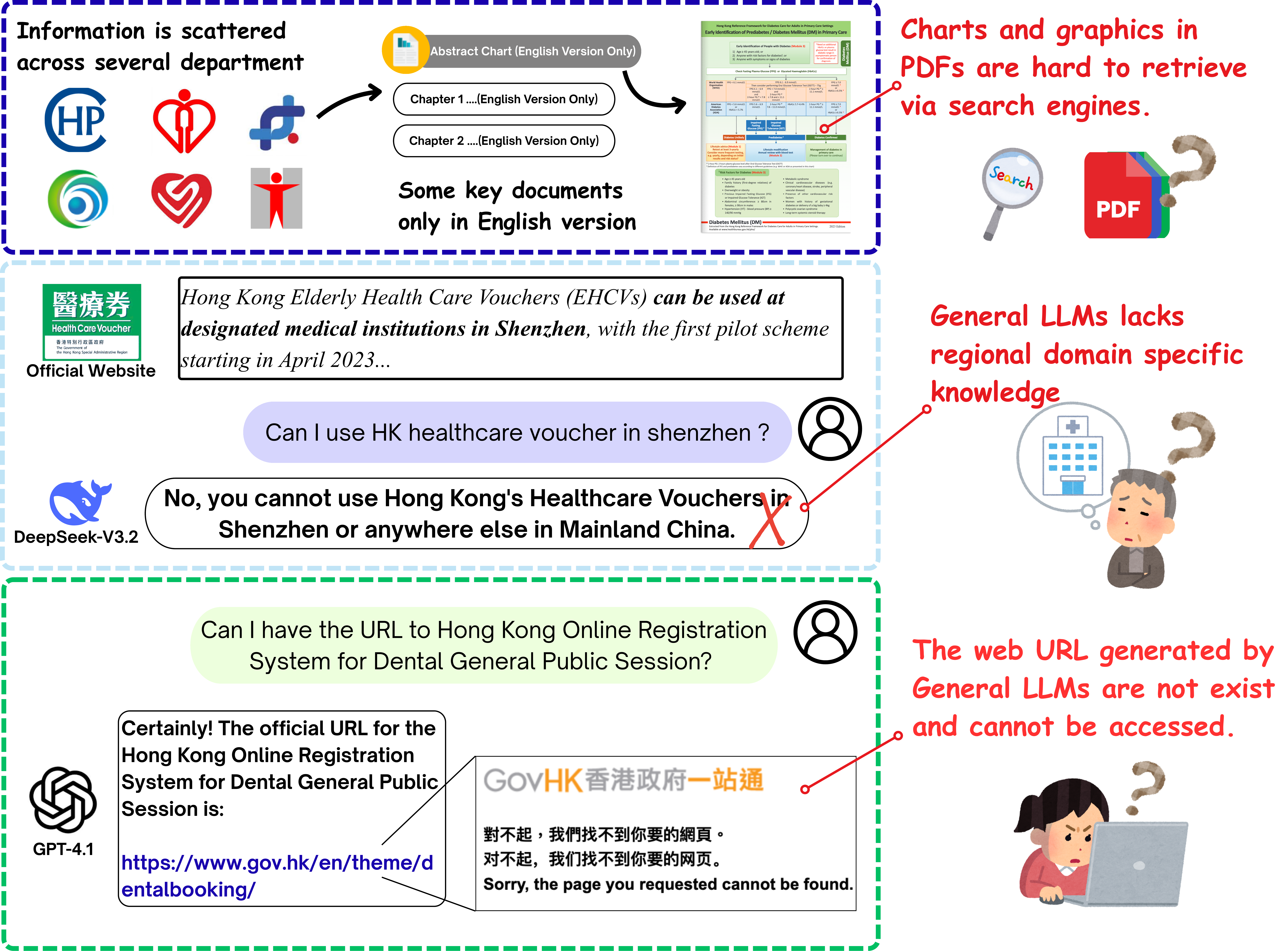}
    \vskip -0.1in
    \caption{The illustration of scattered primary healthcare information and the existing limitations of Large Language Models (LLMs)-based Healthcare Assistant in Hong Kong. }
    \label{HKEP}
    \vskip -0.25in
\end{figure}

With the widespread adoption of Large Language Models (LLMs), there is an increasing trend to use these tools to obtain health information and medical advice ~\cite{Pal2025,fan2024survey,zhou2025hd}.
However, individuals are prone to overtrusting these AI-generated results, 
viewing them as equally effective as professional medical guidance~\cite{doi:10.1056/AIoa2300015,fan2025computational}. 
One primary concern is the tendency toward factual inaccuracies and hallucinations when handling domain-specific and region-specific queries ~\cite{ning2025survey,zhao2024recommender}. 
As illustrated in Figure \ref{HKEP}, the LLM-empowered AI systems might provide incorrect answers about medical services and invalid URLs in Hong Kong. 
Furthermore, these systems often struggle with conversational clarity. When a user asks a fuzzy question such as \textit{"Which exercise is better for my knees?"}, general-purpose LLMs usually fail to clarify crucial context, such as age or medical history, instead offering universal answers that force users into an iterative cycle of questioning. 
Since responses cannot be cross-verified, this also creates significant trust issues for users who cannot verify their validity.

To this end, we focus on the primary healthcare sector and propose a novel \textbf{Pri}mary \textbf{H}ealthcare \textbf{A}ssistant (\textbf{PriHA}) in Hong Kong, where a three-stage pipeline is proposed to build upon our Dual Retrieval-Augmented Generation (DRAG) framework. 
The process begins with an intelligent triage module that precisely captures and clarifies the user’s true intent through multi-turn dialogue. 
A list of clarified sub-queries is then passed to the core DRAG framework, which performs dual-source retrieval. Finally, an enhanced generation module fuses and validates information from both sources, producing comprehensive answers with source citations. Our main contributions can be summarized as follows:
\begin{itemize}[leftmargin=*]

\item We propose a RAG-enhanced LLM framework (\textbf{PriHA}) to address information gaps by optimizing the conversation-retrieval workflow and strategically guiding users toward community resources for self-management in Hong Kong.

\item We introduce the Dual Retrieval-Augmented Generation (DRAG) framework, which resolves conflicts between static and dynamic data sources, thereby eliminating context pollution while providing traceable evidence. 

\item Our comprehensive experiments demonstrate that our proposed method significantly outperforms single-source RAG methods in accuracy and completeness, establishing a responsible AI design paradigm that can be extended to other public-service domains.

\end{itemize}

\section{Related Work}

As one of the most representative AI techniques~\cite{wang2020traffic,fan2019graph,fan2020graph}, LLMs have achieved significant success in the healthcare domain \cite{CHEN2025151,qu2024tokenrec,qu2025diffusion}
 
Current applications range from medical record summarization and patient education to complex diagnostic support \cite{Wang2024}. However, the practical deployment of these systems in high-stakes clinical settings is frequently hampered by their propensity for hallucination, where models generate confident yet factually erroneous information \cite{Asgari2025,Kim2025}. 
Furthermore, due to the static nature of their training corpora, LLMs struggle to incorporate evolving medical knowledge, which leads to potential obsolescence and safety risks \cite{fan2024survey}. Consequently, relying solely on the parametric knowledge of LLMs is increasingly viewed as insufficient for reliable healthcare assistance and necessitates mechanisms that can ground model outputs in verifiable evidence \cite{ning2025survey,fan2024survey}.

To mitigate these limitations, Retrieval-Augmented Generation (RAG) \cite{wang2025knowledge,ning2025retrieval} is a technique that uncovers new possibilities for reliable medical AI by separating knowledge storage from reasoning capabilities. In this framework, models rely on external repositories to retrieve relevant documents and supply them as context for response generation. Compared with fine-tuning, which requires resource-intensive retraining and risks catastrophic forgetting, RAG offers a distinct advantage in adaptability \cite{ning2025survey,wang2025knowledge}. Since medical knowledge and government schemes are dynamic and constantly updated, retrieval-based methods allow for the immediate integration of new information by simply updating the external database \cite{pingua2025medical}. Empirical studies such as those by Xiong et al. \cite{xiong-etal-2024-benchmarking} demonstrate that RAG approach are better than fine-tuned models in handling medical factual queries for source with reference answers. This makes RAG a preferred architectural choice for environments characterized by scattered but comprehensive information, such as primary healthcare.

Despite the success of retrieval systems, developing holistic healthcare assistants for specific populations, such as the elderly, presents unique challenges that remain underexplored. Most existing medical QA chatbots operate on a single-turn or passive-retrieval basis and may fail when users cannot articulate their needs precisely \cite{10.5555/3737916.3738824}. Moreover, relying on a single knowledge source often limits the comprehensiveness of the answer. Emerging research suggests that hybrid approaches combining structured local knowledge with real-time web information are essential for robust performance \cite{jang-etal-2025-medtutor}. 
Our work bridges this gap by proposing a dual-retrieval framework that not only ensures data timeliness but also optimizes the interaction process for clarity and trust.

\section{Method}
The primary healthcare AI assistant system employs a three-stage pipeline: (1) interpreting user intent, (2) retrieving information from static and dynamic sources, and (3) verifying and synthesizing the results. 

\subsection{Primary Healthcare Repository (PHR)}
The foundation of PriHA is to address the fragmentation of healthcare information caused by the dispersion of official guidelines across disparate government websites. We have created a centralized knowledge repository that covers major primary healthcare information in Hong Kong, including clinical standards, community resources, and subsidized schemes, serving as the knowledge base for the RAG process.

One challenge during the pre-processing stage is that files stored in PDF use a fixed layout designed for consistent visual presentation, which is not conducive to machine semantic understanding and structured text extraction. Therefore, source documents are converted into Markdown format to preserve structural integrity while facilitating efficient retrieval. While image content is captured as web links and manually annotated to ensure it can be retrieved later, "source of file" and "updated time" are labeled as metadata for later conflict resolution.

\begin{figure*}
    \centering
    \vskip -0.35in
    \includegraphics[width=0.9\linewidth]{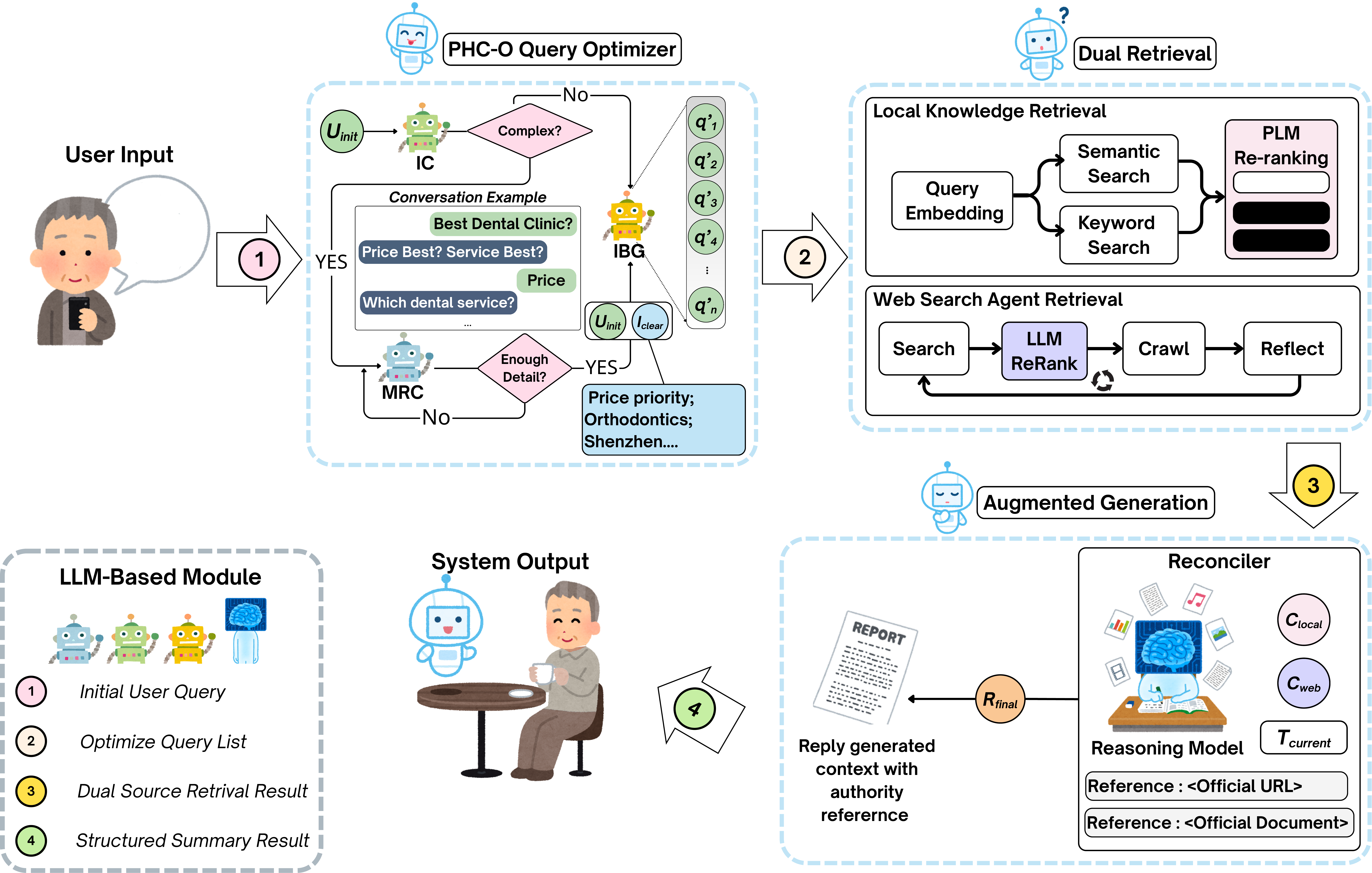}
    \vskip -0.15in
    \caption{Overall framework of the proposed PriHA. The framework processes the initial user input through a query optimizer, yielding a clarified intention and refined sub-queries, then passes to dual retrieval module that fetches and re-ranks content from a local knowledge base and web searchers. Finally, the summarization stage uses a reconciler to synthesize search results into a structured response with proper references.}
    \label{Overall Pipeline}
    \vskip -0.15in
\end{figure*}

\subsection{Primary Healthcare-Oriented Query Optimizer }
General LLMs are not optimized for primary healthcare applications. When presented with a health-related query, these models often generate generic recommendations, such as directing users to the emergency room or hospitals. This response pattern bypasses the local primary care layer.  
To this end,  the optimizer is designed to interpret the user’s initial input ($U_{\text{init}}$) and transform it into a set of PHC-oriented sub-queries ($q'_1, \ldots, q'_n$), aiming to steer the search process toward relevant primary care options such as family doctors and community health centers, rather than defaulting to secondary and tertiary care institutions.

\noindent  \textbf{Intent Classifier (IC)}
Similar to the triage system in an emergency room, where a nurse first confirms the urgency of a patient’s issue, this method uses an LLM-based classifier to process the user’s initial input at the start of the conversation. 
A SIMPLE query is labeled as a direct question with clear intent, such as “\textit{What is the address of Queen Mary Hospital?}” 
and can proceed immediately to the next step. In contrast, a COMPLEX tagged query, such as “\textit{Which clinic has better dental services?}”, is routed to a clarification module. This approach ensures that model  resources are focused on resolving ambiguous queries,
particularly for the elderly users who may struggle to express their needs clearly \cite{LoChan2023} in obtaining desired answers.

\noindent  \textbf{Multi-Round Clarification (MRC)}
When a query is flagged as COMPLEX, the system proactively asks clarifying questions in multiple rounds to understand the user’s intent. For example, when a user mentions “\textit{Where to find a better dental clinic}”, the system helps them clarify the actual meaning of \textit{“better”}, asking whether they are more concerned  about price or something else. The process continues until the user’s intention, background, and priorities are clear, and it is recorded as the clarified user intention ($I_{\text{clear}}$) proceed to next step. This enables the system to perform the retrieval process with a clear understanding of the user’s priorities and background, rather than relying solely on keyword relevance, which might lead to imprecise responses.

\noindent  \textbf{Intent-Based Generalizer (IBG)}
Vanilla generalizers improve coverage simply by paraphrasing words. We propose an Intent-Based Generalizer (IBG) that generates intention-based sub-queries. For example, a query like “\textit{Whether it’s better to see a dentist in Shenzhen or Hong Kong?}” can be refined into searches like  “\textit{Hong Kong dental prices}” and “\textit{Shenzhen dental prices}” after clarification reveals that the user intends to find more affordable services.

Moreover, IBG also incorporates the community-first principle of primary healthcare. For instance, if a user expresses concern about joint pain, the generalizer generates search intents not only for orthopedic guidelines but also for relevant services at district health centers and government-subsidized care schemes. This process produces a list of atomic search queries  ($q'_1, \ldots, q'_n$), to make the retrieval process significantly more comprehensive.

\subsection{Dual-Retrieval Augmented Generation}
Each query $q'_i$  in the generated list is executed in parallel by two independent retrieval components, and their results are then reinforced onto the generative output according to a set of rules.

\noindent  \textbf{Local Knowledge Retrieval (LKR)}
Some official healthcare materials mention the main subject only in the title, without further elaboration in the body. Using a fixed chunk size in such cases can easily lead to incomplete answers due to missing context, but increasing the chunk size will dilute semantic focus.

We adopt a parent-child chunking approach, using two distinct chunk sizes to address the problem. Smaller child chunks are used for retrieval, while larger parent chunks provide context. During the search process, child chunks containing concise, query-relevant details are selected, and their corresponding original parent chunks are forwarded to the system. This method ensures precise retrieval while providing the LLM with sufficient background information during the answer generation step.

Two complementary retrieval methods are employed on the PHR:
\begin{itemize}[leftmargin=*]
    \item \textbf{Keyword Search.} This method operates by matching keyword frequencies and is highly effective for retrieving exact policy terms such as “\textit{Elderly Health Care Voucher}”, but it fails to capture semantic similarity, for example, when users ask about “\textit{medical discounts for seniors}.”
    \item \textbf{Semantic Search.} This method captures semantic similarity and effectively understands implied user intent, but it may lose key proper nouns during vector-space averaging when processing jargon.
\end{itemize}
To leverage the strengths of both approaches, we implement a hybrid retrieval strategy. Candidate chunks from keyword and semantic searches are merged into a unified pool,
followed by a lightweight PLM-based \textbf{BGE-Reranker-V2} applying top-k filtering and score-based selection. The system ensures that only the most relevant contexts are forwarded to the response generation stage.

\noindent  \textbf{Web Search Agent (WSA)}
Although a local knowledge repository provides a reliable information cornerstone, its utility is limited by the latency associated with real-time updates and pre-processing cycles. In practice, information such as service policies and clinic operating hours changes frequently, rendering static data insufficient for a robust system. To overcome this limitation, the framework is equipped with a suite of integrated web search and crawling tools that the LLM-based Web Search Agent (WSA) can invoke dynamically.

The agent operates within an iterative ReAct-style loop. This cycle begins with the formulation of a preliminary plan, followed by a broad web search to obtain an overview of relevant information. Standard web search engines, such as Google Search and Baidu Search, may return links of uncertain credibility when used without carefully designed keyword constraints. To mitigate this risk, we define a safelist that allows the agent to select authoritative URLs from the search results.
To ensure reliability, all search results undergo an LLM-based reranking process. In contrast to traditional PLM-based rerankers, the LLM-based approach offers zero-shot ranking capabilities and the ability to interpret complex, instruction-driven criteria. The reranking rules prioritize official and certified sources to enforce credibility standards.
After reranking, the agent crawls the selected URLs to extract content. The crawler validates the accessibility and authenticity of every link in real time, eliminating broken, fake, or redirected URLs before any content is accepted. The system then evaluates whether the obtained information is sufficient to answer the original query. If gaps or inconsistencies remain, the WSA enters a reflection phase, identifies missing or unreliable data, and triggers another iteration of searching, reranking, and crawling.

Through this iterative process with URL validation via real-time crawling, the agent guarantees that all citations in the final response are verifiable, up-to-date, and directly accessible.

\noindent  \textbf{Augmented Generation (AG)}
At this stage, the system combines information retrieved from the local knowledge base with content acquired by the web-search agent. The challenge is that the simple concatenation of these sources is inadequate, as it may lead to conflicting statements and inconsistent or incoherent outputs. To address this, a systematic reconciliation mechanism is required to integrate heterogeneous evidence in a principled manner.

The Reconciler module is the core component of the AG stage. It jointly evaluates the context retrieved from the local knowledge base ($C_{\text{Local}}$), the web search agent ($C_{\text{Web}}$), and the current user profile ($T_{\text{current}}$). This component operates under predefined conditions that guide decision-making and prioritize information. For instance, official policy documents bearing institutional certification from the Hong Kong government are deemed more reliable than web pages from other administrative sources. Similarly, sources containing outdated content are assigned lower priority. This structured decision procedure prevents the model from presenting unresolved contradictions to the user and ensures coherent synthesis across diverse sources.

We employ a reasoning model to perform conflict resolution and summarization in a single pass. The model integrates comprehensive professional medical guidelines vertically and horizontally with available community resources, such as social welfare programs and complementary schemes. It then constructs a unified representation of the retrieved information and formats it into a report-like response tailored to the end user, emphasizing clarity and readability. This approach ensures that policy and healthcare guidance remain accessible and unambiguous.

Traceability is a central design of our objective. The local knowledge base provides metadata-annotated references, and the web-search agent supplies verifiable URLs obtained via the crawler. These sources are documented in a reference section accompanying the final system output ($R_{\text{final}}$). The capacity for cross-verification through explicit citations enhances the system’s transparency and trustworthiness, thereby increasing its applicability in sensitive, high-stakes domains.

\section{Experiment}

\subsection{Experiment Setting}

\subsubsection{Datasets}
\begin{wrapfigure}{r}{0.5\textwidth}
    \centering
    \vskip -0.15in
    \includegraphics[width=0.9\linewidth]{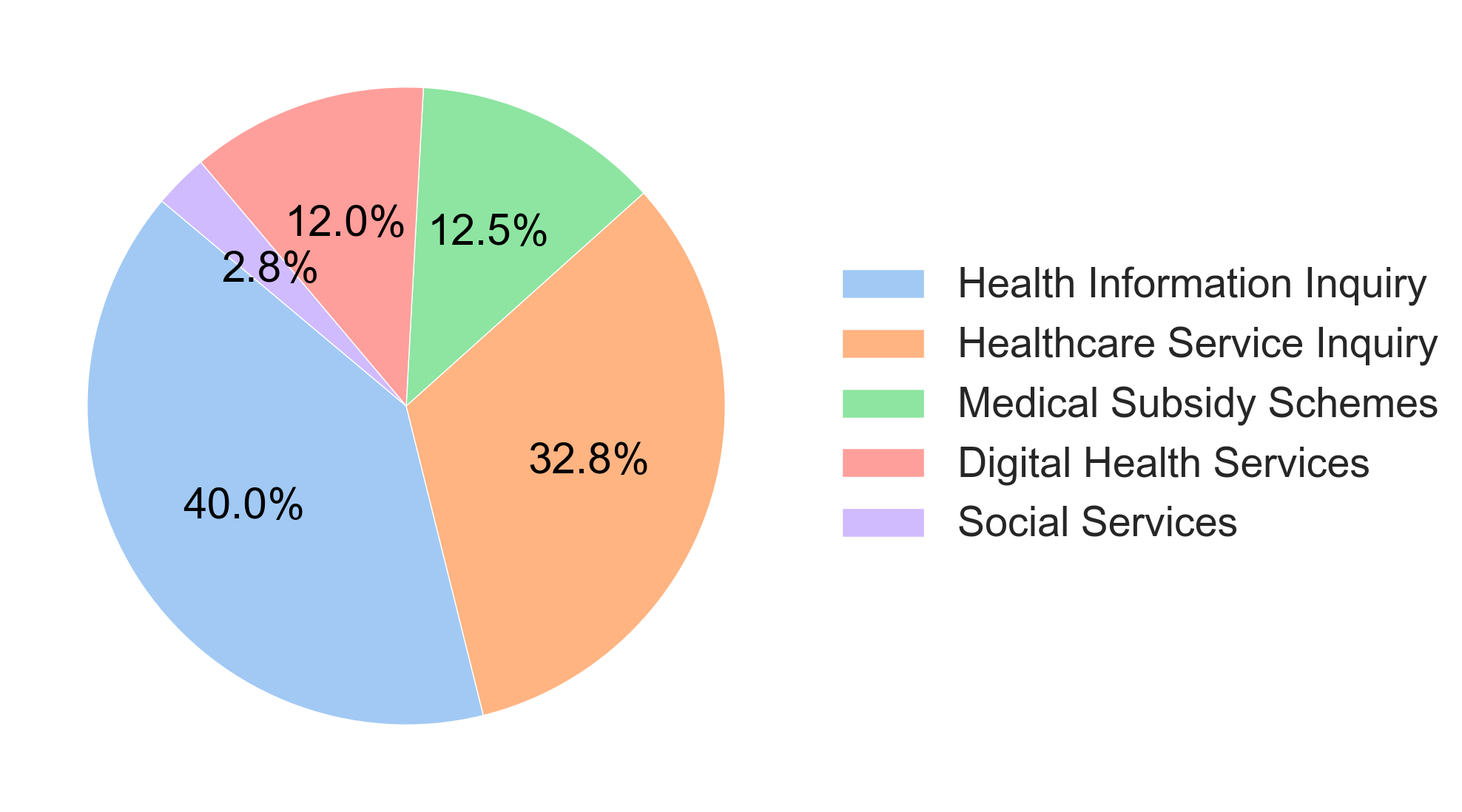}
    \vskip -0.15in
    \caption{Distribution of the HK-PriHCQA dataset by category.}
    \label{fig:dataset_comp}
\end{wrapfigure}

To evaluate the system's performance within the specific context of the Hong Kong medical system, we constructed the Hong Kong Primary Healthcare Question Answering (HK-PriHCQA) dataset. 
This dataset comprises 400 Q\&A pairs derived from official government publications and public health department documents. To ensure ecological validity, questions were developed using two strategies: extracting existing FAQs and paraphrasing topics from informational documents into natural language queries. As shown in Fig. \ref{fig:dataset_comp}, the dataset covers a broad spectrum of domains relevant to primary healthcare.

\noindent  \textbf{Evaluation Metrics}
We employed an "LLM-as-a-judge" methodology~\cite{li2025generation} to capture the nuanced requirements of our target users, prioritizing empathy and clarity alongside accuracy. We defined five key metrics, detailed in Table \ref{tab:evaluation_metrics}, each evaluated on a 6-point Likert scale (0-5).

\begin{table}[h]
    \centering
    \small
    \caption{Evaluation Metrics Definition}
      \scalebox{0.75}{

    \begin{tabular}{l p{9cm}}
        \toprule
        \textbf{Metric} & \textbf{Definition} \\
        \midrule
        Accuracy        & The answer is factually correct and consistent with ground truth. \\
        Completeness& The response covers all aspects of the query and provides the necessary context.\\
        Trustworthiness & The response cites specific, verifiable, and authoritative sources. \\
        Clarity& The response uses clear, simple language, adopts an empathetic tone, and is well-structured and easy to read.\\
        Relevance& The response directly and completely addresses the specific question asked.\\
        \bottomrule
    \end{tabular}
    }
    \label{tab:evaluation_metrics}
    \vskip -0.25in
\end{table}

\noindent  \textbf{Baseline and Ablation}
We compared our proposed framework against three configurations to measure component contributions:

\begin{itemize}[leftmargin=*]
    \item \textbf{Baseline Zero-shot LLM (DeepSeek-V3.2-Exp):} Prompts the model directly without retrieval, quantifying inherent parametric knowledge.
    \item \textbf{Local-Only RAG:} A single-source system using only the PHR knowledge base, evaluating performance in isolation from web search.
    \item \textbf{Web-Only RAG:} Relies solely on a web search engine without the white list and re-ranking process, evaluating the risks of unconstrained information sources.
\end{itemize}

\noindent  \textbf{Limitation}
The query optimizer component, initially designed to handle ambiguous queries, was disabled for this experiment. This ensures a fair methodological comparison by maintaining a straightforward question-answering task structure rather than introducing the inherent variances of multi-round dialogue.

\subsection{Experiment Result}
The evaluation was conducted on a random subset of 80 questions from the HK-PriHCQA dataset. As presented in Table \ref{tab:metric_breakdown}, the proposed DRAG framework outperforms all methods, with the highest overall mean score (4.197). 

\begin{table}[h]
\centering
\vskip -0.15in
\small
\setlength{\tabcolsep}{4pt} 
\caption{Mean scores by metric for each system ($N=80$)}
\label{tab:metric_breakdown}
\vskip -0.15in
  \scalebox{0.75}{
\begin{tabular}{l c c c c c c}
\toprule
\textbf{System} & \textbf{Acc.} & \textbf{Comp.} & \textbf{Trust.} & \textbf{Clar.} & \textbf{Rel.} & \textbf{Avg.} \\
\midrule
Zero-shot LLM  & 3.23 & 3.11 & 3.35 & 4.75 & 4.24 & 3.74 \\
Web-Only RAG & 2.54 & 2.23 & 2.64 & 4.66 & 3.88 & 3.19\\
Local-Only RAG & 3.50 & 3.35 & 3.43 & 4.71 & \textbf{4.34} & 3.87\\
\textbf{DRAG}      & \textbf{3.95} & \textbf{4.03} & \textbf{3.98} & \textbf{4.86} & 4.18 & \textbf{4.20} \\
\bottomrule
\end{tabular}
}
\vskip -0.15in
\end{table}

\begin{figure}[h]
    \centering
    \includegraphics[width=0.88\linewidth]{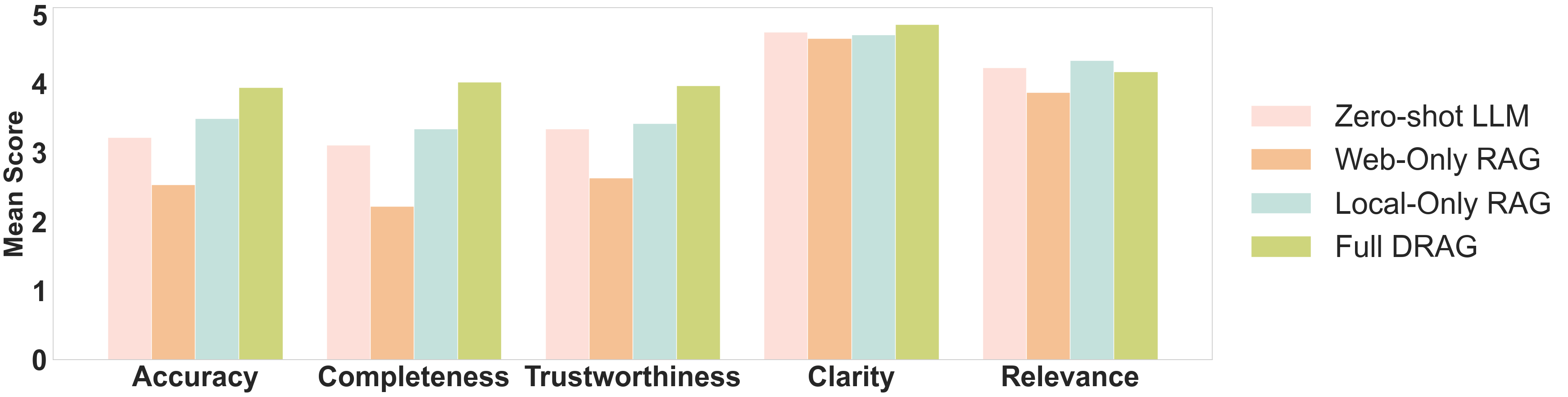}
    \vskip -0.15in
    \caption{Comparative performance of system configurations across five metrics.}
    \vskip -0.15in
    \label{fig:experiment_result}
\end{figure}

The results presented in Figure \ref{fig:experiment_result} reveal a clear trade-off among the ablation configurations. The \textbf{Web-Only RAG} achieved the highest completeness score but recorded the lowest accuracy, demonstrating that unfiltered information volume can degrade factual integrity. Conversely, the \textbf{Local-Only RAG} achieved high relevance owing to its curated content but exhibited the lowest completeness. 
Our \textbf{DRAG framework} effectively addresses these shortcomings by preserving the trustworthiness and relevance of the local knowledge base while leveraging a controlled web-search agent to improve completeness significantly.

\begin{figure}[h]
    \centering
    \vskip -0.15in
    \includegraphics[width=0.9\linewidth]{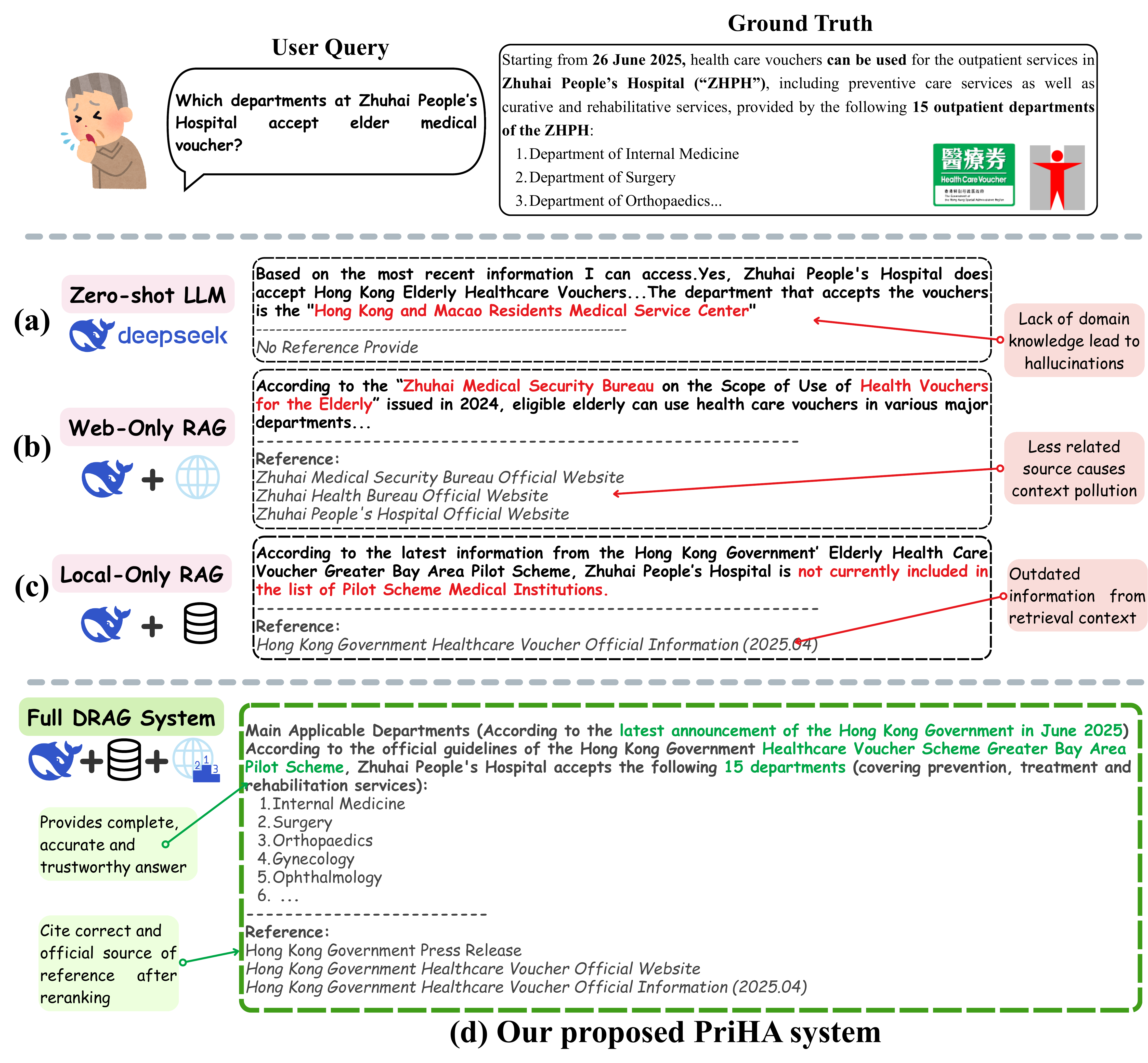}
    \vskip -0.15in
    \caption{Qualitative comparison on a query regarding voucher acceptance at Zhuhai People's Hospital. }
    \vskip -0.15in
    \label{fig:casestudy}
\end{figure}

\subsection{Case Study}
To qualitatively assess the retrieval balance, we analyzed a query regarding the acceptance of elderly health vouchers at Zhuhai People's Hospital (Fig. \ref{fig:casestudy}). This case highlights the critical need for reasoning-based reconciliation. The \textbf{Zero-shot LLM} hallucinated, misinterpreting the query as a social security issue. The \textbf{Web-Only RAG} retrieved web content but introduced context pollution from less-related sources. The \textbf{Local-Only RAG} provided a factual but outdated response, failing to reflect the latest policy extensions.

The \textbf{DRAG framework} was the only method that succeeded. It effectively grounded the query in local knowledge to establish the correct context, then used web retrieval to verify recent, authoritative updates. The Reconciler module prioritized the authoritative, time-valid information over the outdated local entry, producing a coherent and empathetic response. This confirms that our framework can produce contextually reliable and transparent answers even under constrained conditions where static databases lag behind policy changes.

\section{Conclusion}
In this paper, we developed PriHA, a Primary Healthcare AI assistant designed to bridge the information gap in Hong Kong’s healthcare system. Beyond ensuring safety and factual accuracy, our system proactively steers users toward community-based resources. By integrating fragmented government guidelines into a centralized knowledge base, the system goes beyond merely answering explicit user queries; instead, it strategically recommends relevant primary healthcare services when users do not explicitly ask for them. This approach not only provides accurate, traceable information but also subtly reshapes user perception, encouraging a shift from hospital-reliant care to prevention-oriented self-management. Future development will focus on automating knowledge updates and integrating voice interaction to enhance accessibility for the general public further.

\begin{credits}
\subsubsection{\ackname} 
The research described in this paper has been partially supported by the General Research Funds from the Hong Kong Research Grants Council (project no. PolyU 15207322, 15200023, 15206024, and 15224524), 
internal research funds from Hong Kong Polytechnic University (project no. P0042693, P0048625, and P0051361), and Sheertek International (HK) Limited.   
This work was supported by computational resources provided by The Centre for Large AI Models (CLAIM) of The Hong Kong Polytechnic University.
\end{credits}

%
%
%
\bibliographystyle{splncs04}
\bibliography{bib-refs}
\end{document}